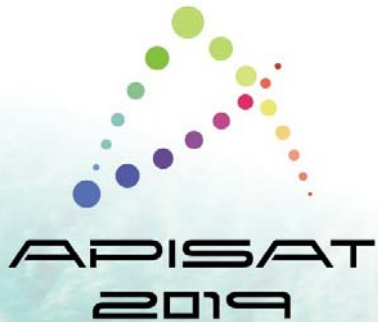
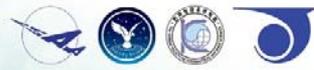

# Weapon-Target Assignment Problem with Interference Constraints using Mixed-Integer Linear Programming

Min-Kyu Shin, Daniel Lee, Han-Lim Choi

Department of Aerospace Engineering, KAIST

Daejeon 305-701, Republic of Korea

[mkshin, dilee, hanlimc]@lics.kaist.ac.kr

In this paper, we propose an approach to formulate the Weapon Target Assignment (WTA) problem with physical and seeker interference constraints which is solvable in Mixed Integer Linear Programming (MILP). To handle the interference constraint which is intractable in continuous time domain, we discretize the time window and generate predicted intercept point (PIP) set. Also, to handle interference constraints to the MILP formulation, the interference tables that explain which pairs of assignments make interference, are made prematurely before the solver executed. The simulation results whether considering interference or not give different optimal solutions with different targets show the effectiveness of our MILP formulation.

## I. INTRODUCTION

As technology develops, both attack and defence systems are available to operate multiple weapon farms and shoot many weapons in short time. Especially, when a large number of weapons attack a small local area, the situation will have dense limitations of time and very hard to shoot. For these reasons, it is very important to get optimal solutions which can intercept the targets as many as possible.

Weapon-Target Assignment (WTA) problem is a well-known problem for assigning several weapons to several targets within limited time window and with set constraints. It is known to be NP-complete problem [1] as various researches have been proposed to get the optimal solution or acceptable solutions quickly. One of the methods to get an optimal solution is using the mixed integer linear programming (MILP) with configuring an objective function and constraints [2]. Many solvers such as CPLEX, GUROBI, and XPRESS are available to solve these kinds of problems in recent times. Although it gives the initial intuition of the problem, the launch delay constraints and interference constraints are missed. Also, the meta-heuristic algorithms to get the benefits of computation speed and acceptable optimality are developed [3-9]. Also, Genetic Algorithm which treats each assignment to genes [4-7] as a general idea is also well- used. The Simulated Annealing (SA) [5, 6] or Tabu Search [8] is used for search solutions. For local changes of the solutions, Ant Colony Optimization (ACO) [7] or Particle Swarm Optimization (PSO) [9] can be accepted. Although [3] uses concepts of physical interference with Evolutionary Strategy, it doesn't consider seeker interference and other meta-heuristic paper [4-9] doesn't consider the interference constraints. Also, Maximal Marginal Rule for solving the algorithm is very useful the solution efficiently [11, 12]. However, if the later solutions are affected by earlier solutions, it cannot guarantee optimal solutions mostly [13]. Therefore, it is not applicable because the forehead assignments affects huge by launch delay and interference in this problem. It is possible to use neural network to solve WTA problems with constraints [10], but still challenging including interference constraints.

In the actual combat situation, it is significant to consider the interference between missiles as it affects the overall system performance. The weapons can collide with each other when they are too close, also the weapon seeker which should capture target, can capture another weapon and head to it when it is in the detection range of the radar. Thus, this paper treats two kinds of interference, physical interference and seeker interference, and proposes a mixed-integer linear program formulation. While it focuses on maximizing number of intercept of targets to minimize damage of assets and interference, the interference conditions are checked more strictly. To avoid physical interference between two missiles, the weapon should be the maximum radius of the missile away. For treating seeker interference constraints, the weapon should not be in detection angle of other weapons' seeker. For these interference constraints, the weapon-target assignment and launch time are carefully considered. Also, not to make interference condition too much, the all interference conditions should be checked preliminarily.

In this paper, to avoid these potential interference problems, we reformed the WTA problem into assigning weapons to predicted intercept points (PIP), not assigning weapons to targets additional constraints. It needs pre-processing for making predicted intercept points. Previously, there are intractably many possibilities for continuous launch time, but discretised and limitary number of intercept points in this reformed formulations, and it makes to find an optimal solution.

Details of the mixed-integer linear program formulation are validated in the section II. Because this paper concentrates on the avoidance of interference, various combat situations, from simple combats to complex combats, are applied to simulation scenarios. The simulation results show that the model effectively intercept missile and make fewer number of interference against existing mixed-integer linear program model.

## II. PROBLEM FORMULATION

In this section a WTA problem with the interference constraints is formulated. A standard WTA problem is formulated solvable in MILP and the interference constraint designed in a linear form are considered.

### A. WTA Problem Formulation

Basically, the final goal of the WTA problems is to find the set of pairs of weapon farm, target and launch time $(w, t, \tau)$ that optimize the cost function, with satisfying the accompanied constraints. MILP formulation of the WTA problem which maximize number of assigning weapons to targets as many as possible, is defined as

$$\min J = \sum_{t=1}^{|T|}\left[\log(v_0(t)) + \prod_{w=1}^{|W|}\prod_{m=1}^{|M_w|}(\theta_{w,t}(m) \times \log(1 - \gamma_{w,t}))\right] + \max(\log(v(t))) \quad (1)$$

In Equation (1), $v_0(t)$ represents the initial value of target $t$ and $\theta_{w,t}(m)$ is a binary variable which represents whether the weapon $m$ of the weapon farm $w$ is assigned to target $t$ or not. If it is assigned, its value is equal to 1, otherwise it is 0. $\gamma_{w,t}$ is the destruction probability of the target due to weapons from the weapon farm. The first term of Equation (1) means the remained total value of targets and the second term represents the maximum value of the targets, after assignment. Considering the case that the initial value of the targets are all identical and decrease constantly, the objective function with only summation of remained targets can give the solution which assigns few targets to many weapons. To prevent that, the second term after the summation term is induced to make the assignment spread out to several targets. To analyse more simply, $v_0(t)$ are all identical to $e^2$ for every targets, and $\gamma_{w,t} = 1-e^{-1}$ for all pairs of weapon and target. This is a trick to make $\log(v_0(t))$ and objective remains integer. When only one weapon is assigned to the target, the target value becomes 1, and two weapons are assigned to the same target, its value becomes 0.

The constraints of WTA include capability constraints, resource constraints, and strategy constraints, and constraints due to processing time.

$$\sum_{t \in T} \theta_{w,t}(m) \leq 1, \quad \forall\, w \in W, m \in M_w \tag{2}$$

Equation (2) states the weapon assignment constraints. Each of weapons goes to single target, and is impossible to hit multiple targets.

$$\sum_{t \in T} \sum_{m \in M_w} \theta_{w,t}(m) \leq |M_w|, \quad \forall\, w \in W \tag{3}$$

Equation (3) states the weapon number or resource constraints. Each weapon farm cannot shoot weapons over the number of missiles that the weapon farm has. Each weapon farms would have limited number of weapons and the weapon farm can shoot the weapons until the weapon farm use up all weapons.

$$|\tau_w(m_1) - \tau_w(m_2)| \geq \tau_w^d, \quad \forall\, w \in W,\ m_1, m_2 \in M_w, m_1 \neq m_2 \tag{4}$$

Eq. (4) states the launch time delay constraints. All weapon farms cannot prepare to shoot two weapons for very short time. The preparation time for shooting another weapon is needed. In this paper, for all weapon farms, we fix $\tau_w^d$ to 1 second which is minimum delay for two weapons from same targets.

$$\sum_{w \in W} \sum_{m \in M_w} \theta_{w,t}(m) \leq n_t, \quad \forall\, t \in T \tag{5}$$

Eq. (5) states the upper bound of maximum assignments for each target. This is a strategy for not too many weapons assigned to a single target. If not, it can make waste of weapons and lead to failure of other possible assignments. Therefore, limiting number of assignment to a single target, $n_t$ can give more reasonable results. In this work, we fix $n_t$ to 2.

*B. Interference Constraint*

While maneuvring the missiles, two types of interference are considered. First, the weapons have a possibility to collide each other because the trajectories of defensive weapons over time are overlapped. We defined it as a physical interference and to avoid it, every pair of missiles should retain a certain distance, considering the missiles' radius and safety margins. The other type of interference is the seeker interference. Each weapon has a seeker to help the terminal guidance to intercept exactly. If other weapons are in the detection range, the weapon can recognize our weapons as targets. Each of the interference is represented in Figure 1.

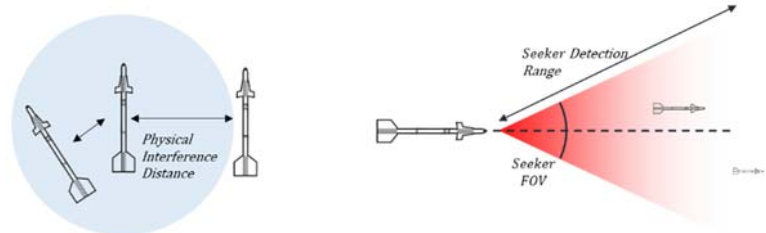

**Figure. 1 Interference definition by type - left : physical, right : seeker**

To provide a constraint to the WTA problem the interference should be checked for every pair of trajectories as

$$\theta_{t1,w1}(m_1) + \theta_{t2,w2}(m_2) \leq 1 \quad \forall (\theta_{t1,w1}(m_1), \theta_{t2,w2}(m_2)) \in I \qquad (6)$$

Where *I* is the interference information of every possible assignment pairs. Then Equation (6) states that if two assignments make interference, at least one assignment should be eliminated to prevent the interference.

In conclusion, the final goal of this paper is to optimize the cost function (Equation (1)), with satisfying the technical, strategical constraints (Equations (2-5)), and interference constraints (Equation (6)).

## III. METHODOLOGY

### A. Discretization & PIP set Genetation

To solve the WTA problem defined in Section II the interference information, *I*, in Equation (6) should be constructed prior. Since *I* should be determined by the assignment results, which weapon farms are assigned to which targets at which time, it should provide information for every possible assignment. Especially, if we handle this problem in continues time domain, the size of *I* will be infinite and intractable. To overcome this, we discretise the time by converting the target trajectory to a set of PIPs (predicted impact point), within the minimum and maximum defence range as depicted in Figure 2. To avoid interference assignments, the target trajectories should be divided by reasonable number to make PIP set that can check physical interference.

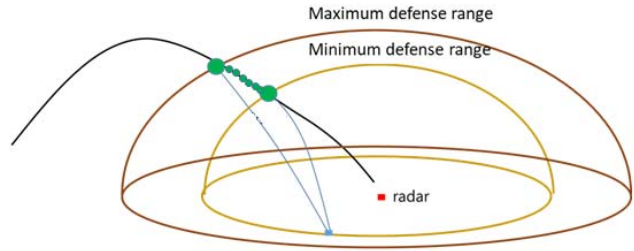

**Figure. 2 Discretization and PIP set**

This approach has advantages of that the problem turn to be tractable and the PIP set can be directly applied to the WTA problem formulation.

$$\sum_{t \in T} \sum_{i=1}^{|PIP|} \theta_{w,t}(m,i) \leq 1, \quad \forall\, w \in W, m \in M_w \qquad (7)$$

Equation (7) states the additional constraint that a weapon should be assigned to a single PIP point from the PIP set of the target.

### B. Construction of Interference Information, I

To construct *I*, every possible pair of assignments should be compared for the overall trajectories. For that we develop a guided weapon dynamics simulation method with the dynamics as

$$\dot{x} = V_x, \quad \dot{y} = V_y, \quad \dot{z} = V_z \qquad (8)$$

Where $x, y, z$ are the states of the weapon and $V_x, V_y, V_z$ are the velocity of each state direction which are given as

$$\dot{V}_x = \frac{(T_x - D)(cos\gamma cos\psi) - (T_x tan\beta + Y)sin\psi + (-T_x tan\alpha - N)(sin\gamma cos\psi)}{m}$$
$$\dot{V}_y = \frac{(T_x - D)(cos\gamma sin\psi) - (T_x tan\beta + Y)cos\psi + (-T_x tan\alpha - N)(sin\gamma sin\psi)}{m} \quad (9)$$
$$\dot{V}_z = \frac{-(T_x - D)sin\gamma + (-T_x tan\alpha - N)cos\gamma}{m} + g$$
$$T_x = T/(\sqrt{1 + \tan^2\alpha + \tan^2\beta})$$

Here $T_x$ denotes the thrust to the $x$ direction and $\alpha, \beta$ are the angle of attack and side slip angle, respectively and $g, m$ are the acceleration of gravity and the mass of weapon. $\gamma, \psi$ are the Euler angles and $D, Y, N$ are the aerodynamic forces

$$D = C_D Q S_{ref}, \quad Y = C_Y Q S_{ref}, \quad N = C_N Q S_{ref} \quad (10)$$

Where $C_D, C_Y, C_N$ are the aerodynamic coefficients and $Q$ and $S_{ref}$ are the dynamic pressure and reference surface area.

The weapon has a short boost phase and then guided to the target or PIP with the three dimensional pure proportional navigation guidance laws given as

$$\Omega = \frac{r_M^T \times V_M^T}{|r_M^T|^2}$$
$$a_{PNG} = N|V_M|\Omega \times \frac{V_M}{|V_M|} = N\Omega \times V_M - g_0 cos\gamma \quad (11)$$

Where $\Omega$ is the line of sight rate and $r_M^T, V_M^T$ are the relative position and velocity vector of the target and the weapon. $V_M$ is the velocity of the weapon. The last term of Eq. (11) denotes the acceleration to compensate the gravity effect.

Based on the simulation method, we can obtain the guided weapon trajectories, and can check numerically whether they violate the interference defined in Figure.1. For physical interference as

$$|x_2(t) - x_1(t)| < D_{PIF} \quad \forall t \in [t_{01}, t_{go1}] \cap [t_{02}, t_{go2}] \quad (12)$$

Where $x_1, x_2$ denotes the state vectors for any pair of weapon trajectories, $D_{PIF}$ is the physical interference distance and $t_0, t_{go}$ are the launch time and terminal time for each weapon with the proper subscripts respectively.

For seeker interference we should check both of the distance condition and the angle condition as

$$|x_2(t) - x_1(t)| < D_{SIF} \quad \forall t \in [t_{01}, t_{go1}] \cap [t_{02}, t_{go2}]$$
$$cos^{-1}(\frac{r \cdot v}{||r|| \times ||v||}) < \sigma_{FOV} \quad \forall t \in [[t_{01}, t_{go1}] \cap [t_{02}, t_{go2}] \quad (13)$$

Where the first line in Equation (13) denotes the seeker distance condition with $D_{SIF}$, the seeker interference distance, and the second line denotes the seeker angle condition where $\sigma_{FOV}$ is the seeker field of view angle and $r = x_2 - x_1$, $v = \dot{x}_2 - \dot{x}_1$ are the relative position vector and the relative velocity vector of the weapon pair, respectively. After checking Equation (12) and (13) we can update the interference information, $I$.

In this paper we use a two dimensional table form on $I$ for a pair of weapon farms, where each element represents a certain pair of assignment to two PIP points.

## IV. Simulations and Results

In this section, the simulation settings and the results of MILP solver are described. Scenarios of defencing multiple targets with two weapon farms are considered, as two weapon farms are the basic setting for our problem. The proposed approach has the scalability for the increasing number of weapon farms. Two cases are simulated to check the effects of interference and compare the optimality and computation time with respect to scale of the problem.

### A. Simulation Setting

To set the simulation, we consider 6 enemy launchers located in a W shape formation with the increasing numbering from left to right. Each enemy launcher has the ability of launching a target with launch time delay of 2 seconds. With the purpose of comparing how the interference and assignment are affected by the dense of attack two cases are considered. For case1, we assume every enemy launcher shoots a single target simultaneously. And for case2, every enemy launcher launches another target in 2 seconds. The enemy launchers shoot the target to the origin where the asset for protection is assumed to be located. The altitude of interception is intended to be from 4km to 5km, then each target trajectory is discretised to 30 PIP points considering the physical interference distance between each PIP points. Figure 3 shows the simulation setting and the PIP sets, left figure for case1 and right figure for case2. The PIP sets with red color depict the second set of targets which are launched later than the blue ones. Two defensive launchers are assumed to be located in +250m and -250m on the *x* direction. We named first launcher at (250, 0, 0), and second launcher at (-250, 0, 0) for later explanation.

For a realistic combat scenario, we assumed each target is disturb by external forces to form the predicted to hit points as depicted in Figure 4 where $t_{i,j}$ represents the $j^{th}$ target form the $i^{th}$ enemy launcher.

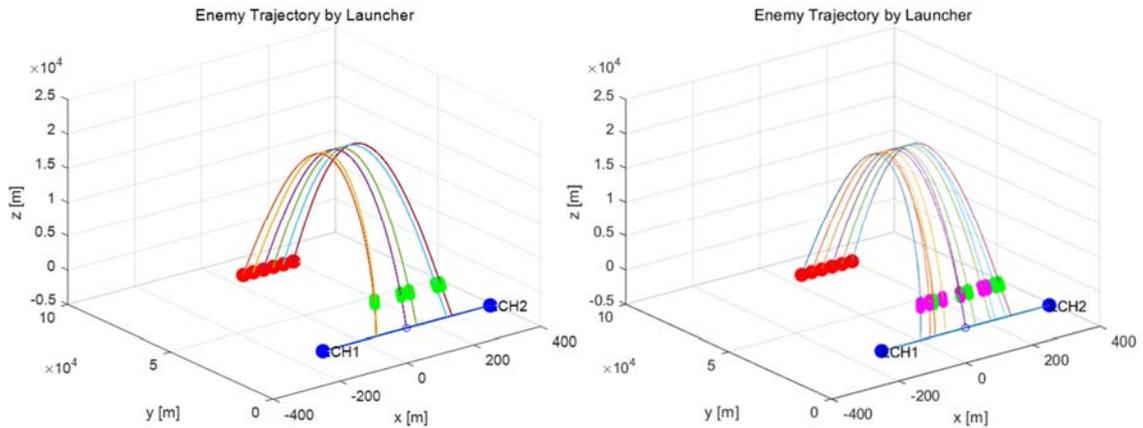

**Figure. 3 Simulation setting and the PIP sets for each case**

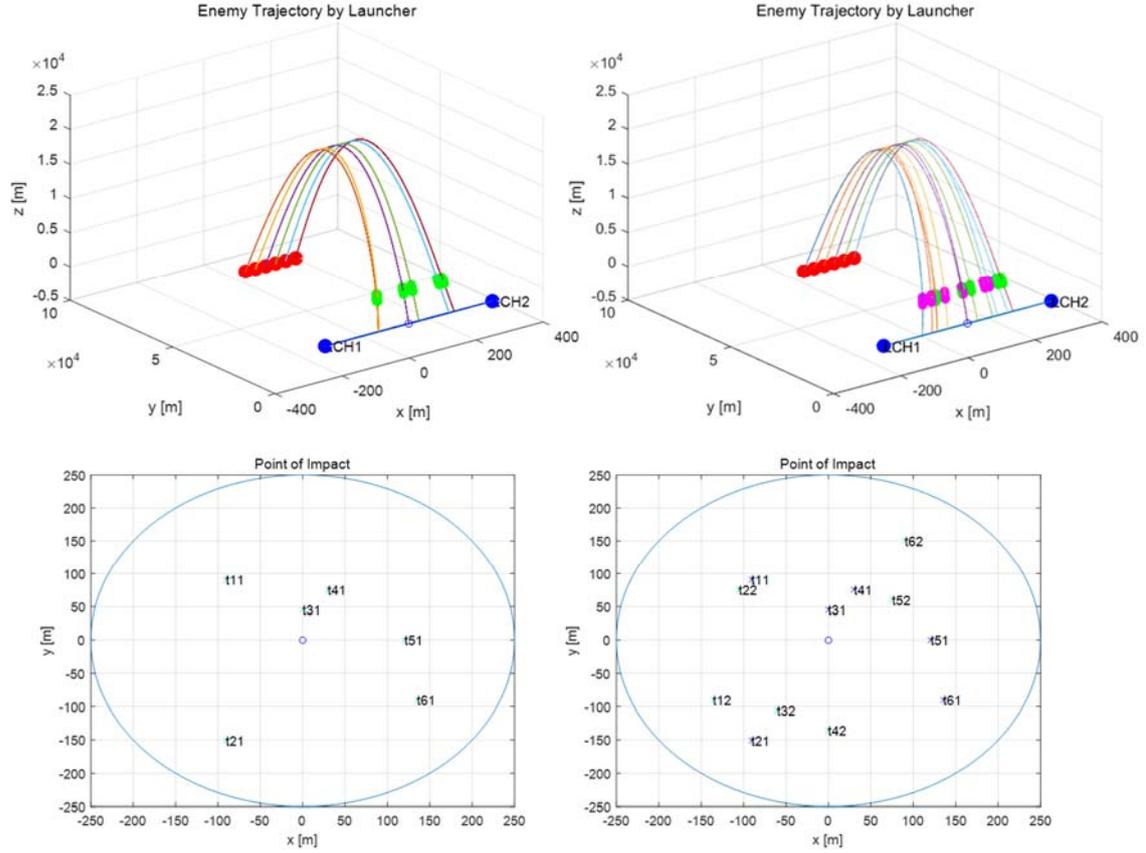

**Figure. 4 Predicted impact points of targets for each case**

To defence the targets, the weapon farms are assumed to have sufficient number of weapons. The launch time delay is 1 second and the maximum number of possible assigned weapon for target is assumed to two. The launch angle for the weapons are assumed to be fixed as $\gamma_0 = 60°, \psi_0 = 0°$.

## B. Inference Information Results

Figure 5 and Figure 6 show the interference information $I$ for each case, respectively. The results are divided for each type of interference for convenience. The parameters for interference are $D_{PIF} = 50m, D_{SIF} = 2000m$ and $\sigma_{FOV} = \pm 8°$. To construct $I$, targets are numbered by the increasing order of enemy launcher number and the group of shoot, for case 2 for example, $[t_{11}(\text{PIPset}), t_{21}(\text{PIPset}), \dots, t_{62}(\text{PIPset})]$.

For case 1, 6 targets are discretised to 180 PIP points and shows 300 and 16243 pairs occur physical and seeker interference respectively, which are 1% and 50% of the total pair number. For case 2, 12 targets are discretised to 360 PIP points and the number of pairs occur physical and seeker interference are 791 and 54319 respectively, which are 0.6% and 42% of the total pair number.

For the detailed characteristics in physical interference in Figure 5, for example, the results can be categorised in two groups with the assignment results $(w_1(t), w_2(t))$, where $w_k(t)$ denotes the assigned target set $t$ for $k^{th}$ weapon farm, $w$. First group, ([1], [3, 4]) and ([3], [4, 5]) occurs physical interference by intersection of the trajectories where each weapon farm is assigned to the target which are closer to the other weapon farm. For the other group, ([4], [3]), occurs physical interference where the assignment has no intersection but the assigned targets are close with similar distance far from each weapon farm. Except the two groups addressed, no physical interference occurred.

In contrast, it is hard to be categorized the cases seeker interference occurs.

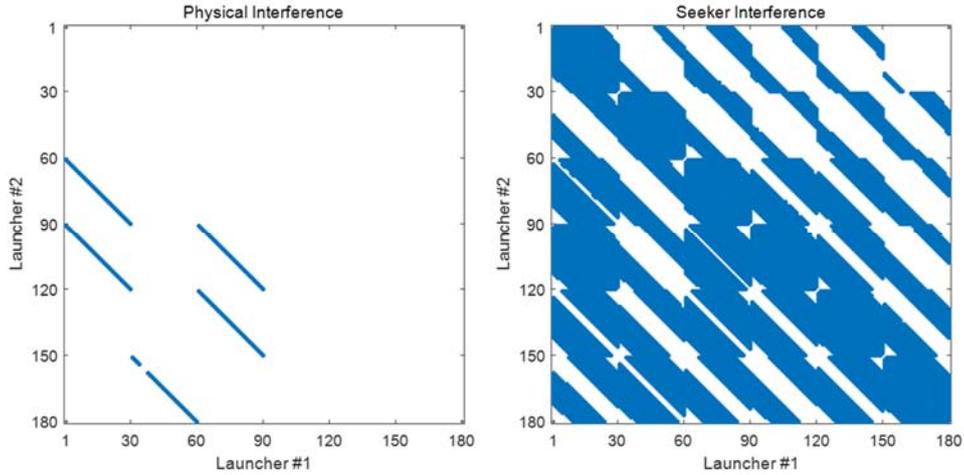
Figure. 5 Interference information results for case1– left : physical, right : seeker

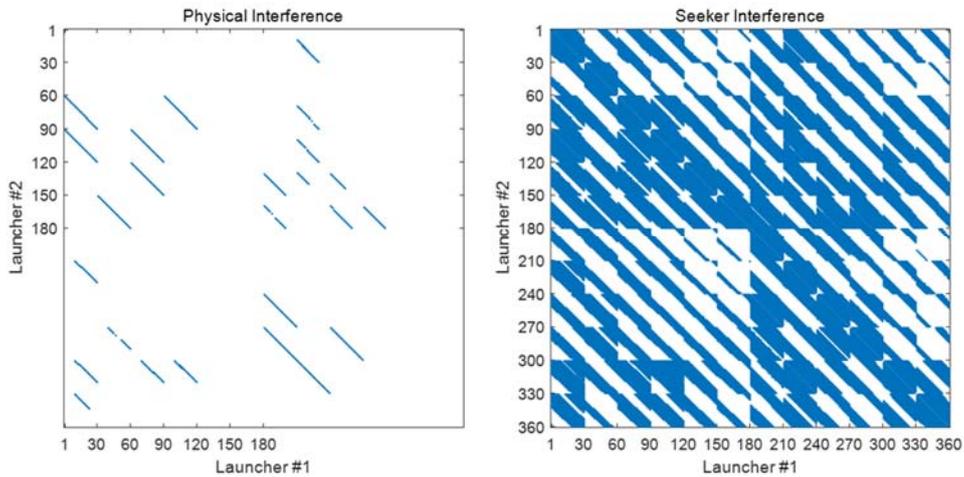
Figure. 6 Interference information results for case2– left : physical, right : seeker

## C. Assignment Results

Figure 7 shows the results with MILP solver. To compare the effects of interference, we simulated fout cases with 2 scenarios with different number of targets whether it considers interferences and not, while remaining the other settings same. Table 1 shows the final objective values when the assignments end. It takes 0.1 seconds and 2.4 seconds without interference constraints for 6 targets and 12 targets respectively. However, it takes 3 hours and over 8 hours with interference constraints for 6 targets and 12 targets respectively. It finds objective solution less than 100 seconds and 15 mins but it takes too long time to be confident that it is optimal. So we configured lower bound as 11 for 12 targets with interference constraints and get the solution. But it still gives a reasonable solution which is similar to the tendency of other results.

From Figure 7(a) and 7(b) shows the result of scenario with 6 targets with and without interference. Figure 7(a) shows that it assigns 10 weapons to the targets. Because at most 2 weapons can be assigned to each target, totally 12 weapons are possible to be assigned in a spatial scenario. However, launch delay is 1 second and the time window for each pair of the launcher and target is less than 6 seconds. Therefore, at most 10 weapons can be assigned in our scenario. For this, remained target value is 2, maximum of remained target is 1 final objective value is 3 finally. However, 2 more weapons miss the targets in the scenario with inteference constraints(Figure 7(b)). Remained target value is 4, and maximum of remained target is 1. So the final objective value is 5.

Figure 7(c) and 7(d) shows scenario with 12 targets with interference and without interference constraints respectively. Without interference constraints, it assigns 16 weapons and miss 8 weapons to targets. With interference constraints, it assigns 11 weapons and 13 weapons miss to targets. In the scenarios with 12 targets, results which interference is not considered, all targets are assigned to by at least one weapon. Otherwise, if the interference is considered, some targets are missed. So the maximum values of remained target, max(log(v(t)) are 2 in this cast. Also still, solution without interference intercepts three more weapons.

For overall results, the objective function well assigns the weapon spreadily. The solutions with interference is not similar with the solutions without interference. Specifically, in the interference condition, the weapon tends to assign the target which is more close to the targets so that the trajectories from weapon to target is not overlapped. As a result, it assigns more weapons when not considering interference, it can fail in the real dense combats. While it takes much time due to increasing the dimension of the problem, it gives the optimal solutions which avoids interference and doesn't fail to intercept.

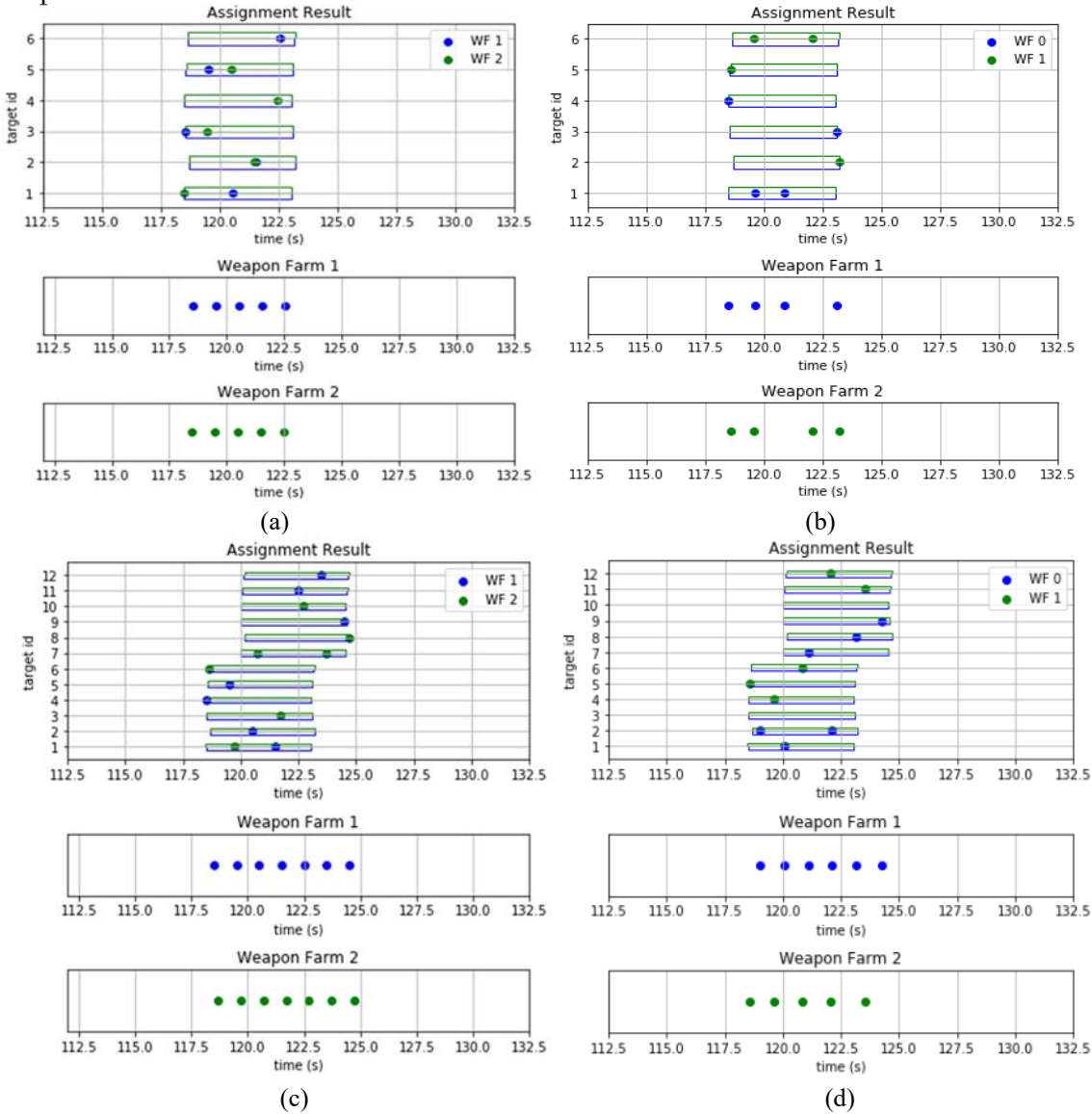

(a) (b) (c) (d)

**Figure. 7 Resulting assignments of the scenario with 6 targets (a) without interference constraints, (b) with interference constraints. Resulting assignments of scenario with 12 targets (c) without interference constraints, (d) with interference constraints. Horizontal lines represent possible launch time from the weapon farm to the target each. And colored dots show launch time of the weapon farms.**

**Table. 1 Assignment Result for MILP with/without Interference**

| Scenario | Number of weapons assigned | Objective value |
|---|---|---|
| 6 targets without interference | 10 | 3 |
| 6 targets with interference | 8 | 5 |
| 12 targets without interference | 14 | 11 |
| 12 targets with interference | 12 | 14 |

## V. Conclusion

In this work, we define a MILP formulation of WTA problem that can handle technical, strategical constraints and interference constraint. With the purpose of evaluate the influence of the combat density two scenarios with different number of targets are performed for both cases with and without interference. The results of interference information, I, and the assignment are analysed. For the interference, the physical interference cases are categorised to two groups while the seeker interference cases are hard to analyse. For the assignment results, the computational time and the performance of the proposed approach are evaluated. The results show that the proposed approach can handle the WTA problems with the interference constraints properly and while considering interference, the solution tends to assign the weapons to closer targets rather than far targets. Although the running time is impractical as weapons and targets increase, it expects to give closely optimal solution to compare with results of other algorithm.

## Acknowledgement

This work was supported by Agency for Defense Development (under contract #UD180017CD)